\documentclass[preprint]{revtex4}
\usepackage{epsfig}
\usepackage{hyperref}
\usepackage{bm}
\usepackage{graphicx}
\begin{document}
\title{Quantum Cosmology Aspects Of D$3$ Branes and Tachyon Dynamics}
\author{Anastasios Psinas}
\affiliation{Department of Physics,\\ Northeastern University \\
Boston, MA 02115-5000, USA} \email{psinas.a@neu.edu}
\begin{abstract}

We investigate aspects of quantum cosmology in relation to string
cosmology systems that are described in terms of the
Dirac-Born-Infeld action. Using the Silverstein-Tong model, we
analyze the Wheeler-DeWitt equation for the rolling scalar and
gravity as well for $R\times{S^3}$ universe, by obtaining the wave
functions for all dynamical degrees of freedom of the system. We
show, that in some cases one can construct a time dependent
version of the Wheeler-DeWitt (WDW) equation for the moduli field
$\phi$. We also explore in detail the minisuperspace description
of the rolling tachyon when non-minimal gravity tachyon couplings
are inserted into the tachyon action.

\end{abstract}
\maketitle

\section{Introduction}

The quest for a fundamental theory which could bring gravity and
quantum phenomena in the same underlying theory, is one of the
driving forces of contemporary theoretical physics. Cosmology, as
a significant laboratory of fundamental physics, can uncover some
aspects of quantum gravity within or even outside the scope of
string theory. One of the major cornerstones in this direction, is
the introduction of minisuperspace formalism for cosmological
backgrounds known as the Wheeler-DeWitt equation
\cite{DeWitt:1967yk}. The crux of this approach is, that the
universe exhibits a quantum mechanical behavior so one can ascribe
a wave function for its dynamics. This idea mainly tells us, that
we can have a sort of a quantum picture for a system that includes
gravity in itself, like our universe. Thus, any information that
we can extract is based upon the exact knowledge of its Hilbert
space that comes as a solution of
\begin{equation}
\mathcal {H}\psi=0 \label{1}
\end{equation}
However, one observes that the right hand side of the equation is
identically equal to zero, which can be seen as a significant
deviation from ordinary Schrodinger's equation. This particular
issue, known as the ``problem of time in quantum gravity'' has
been considered so far a stumbling block preventing us from a
deeper understanding of this formalism. The mainstream idea to
justify the form of Eq.~(\ref{1}) is the belief that there is no
global time in quantum gravity (for an update of this issue see
\cite{Shestakova:2004vr}). Nevertheless, Sen \cite{Sen:2002qa}
based on a former result \cite{Brown:1994py}, showed quite
recently, that in the context of string theory the tachyon field
can be identified with time by writing down a time dependent
version of the Wheeler-DeWitt equation. We will come back and
analyze this advancement later on.

Attempting to solve Eq.~(\ref{1}) requires taking into account
appropriate boundary conditions for the wave function which leads
to different interpretations for the history of the universe.
Among a large number of models that have been proposed, the one
that are mostly discussed and debated are the Hartle-Hawking wave
function \cite{Hartle:1983ai}, the Linde wave function
\cite{Linde:1983mx}, and the tunnelling wave function
\cite{Vilenkin:1984wp}. While we are not going to discuss the main
features of every of them, it is important to say, that these
models not only differ in the techniques utilized to determine the
probability of the universe being in a given state but also differ
through the boundary conditions applied on $\psi$. For instance,
in the tunnelling approach we may even have the emergence of the
universe 'out of nothing' (from zero to a non zero scale factor).
A more detailed account on the intricacies of these models can be
found in \cite{Vilenkin:1998rp}.

Some of the successes of quantum cosmology programme such as
attacking the initial singularity problem, provide some clues for
the inflationary era, its evolution or even its flatness
\cite{Hawking:1985bk} and have drawn the attention of string
theorists. Given the fact that all known string theories contain
apart from gravity a significant number of extra fields such the
dilaton, antisymmetric tensors, RR fields and so on in their
spectrum, one expects that a stringy minisuperspace formalism
would be quite cumbersome. This complexity is manifested in the
wave function which now depends not only on the three-geometries
(in the case of four dimensions) but also on the extra fields that
enter in the string spectrum. As a consequence, the apparent
enlargement of the Hilbert space has been observed and
investigated even in simple scalar field theories coupled to
gravity \cite{Hartle:1983ai,Hawking:1984iv}. As an example, in
\cite{Okada:1985xx} Wheeler-DeWitt equation is solved for
dilatonic gravity, while in \cite{Wu:1984hv} the same equation is
utilized inside the context of Kaluza-Klein cosmologies. In terms
of the gauge gravity correspondence similar important works have
been done along the same lines \cite{Anchordoqui:2000du}. Also, in
\cite{Freedman:2004xg} the Wheeler-DeWitt approach was implemented
in Newtonian cosmology. Various implementations of the
minisuperspace formalism inside the scope of quantum cosmology and
string theory have also been worked on
\cite{Visser:1990wi,Bertolami:1996pc,Gasperini:1997uh,Davis:1999uw,
Rubakov:1999qk,Wiltshire:1995vk,Carlip:2001wq,DaCunha:2003fm,
Firouzjahi:2004mx,Shestakova:2004vr,Kobakhidze:2004gm,
Ooguri:2005vr,Mersini-Houghton:2005im,Sarangi:2005cs,
McInnes:2005su,Hartle:2005ie,Gabadadze:2005ch}.

The progress made
\cite{Sen:1998sm,Sen:2002nu,Sen:2002in,Sen:2002an,Garousi:2000tr,Bergshoeff:2000dq,Kluson:2000iy}
in tachyon condensation, has implications in cosmology. In view of
the work done on non-BPS D-branes a concrete realization on
unstable universes was first introduced in \cite{Mazumdar:2001mm}
followed by \cite{Choudhury:2002xu}. One of the interesting
implications of the introduction of the DBI tachyon action in
cosmological backgrounds, is what has been described as the open
string completeness conjecture \cite{Sen:2002nu,Sen:2002in}.
Basically, the algebraic structure of the system of the FRW and
tachyon equations allows one to consider vanishing velocities for
both the tachyon and the scale factor of the universe at $t=0$.
Also, in \cite{Cremades:2005ir} one finds an implementation of
this hypothesis in warped tachyonic inflation within the KKLT
framework. Further discussions on this conjecture and on tachyon
cosmology can be found in,
\cite{Sen:2004nf,Gibbons:2003gb,Panigrahi:2004qr,Ghodsi:2004wn,Singh:2005qa,Yang:2005rw}
and references therein.

 The advent of AdS/CFT correspondence \cite{Maldacena:1997re},
 has important implications for cosmology.
 One of the important aspects of this duality, is that
 an upper limit on the speed of the moduli fields can be imposed
 \cite{Kabat:1999yq} since causality must be respected. In fact,
 it is known that this motion can be well envisioned in terms of a
 D3-brane moving towards the horizon. Based on that, Silverstein
 and Tong \cite{Silverstein:2003hf} investigated the motion of a
 scalar field coupled to gravity. Indeed in this system, the action
 for the dynamics of the inflaton resembles the DBI action
 \cite{Kutasov:2004dj}. A similar cosmological construction in
 $NS5$-brane backgrounds was investigated by Yavartanoo
 \cite{Yavartanoo:2004wb}. In a different cosmological
 implementation of Ads/CFT duality \cite{Kehagias:1999vr},
stages of cosmological expansions and contractions can occur when
D$3$-branes move along geodesics on a higher dimensional ambient
spacetime. This construction can be utilized for unstable D branes
\cite{Jeong:2005py}, while quite recently in \cite{Psinas:2005ja}
it is shown, that the inclusion of gauge field-tachyon coupling on
the probe brane regulates the cosmological expansion at early
times.

In this paper, some aspects of tachyon physics are presented from
a cosmology point of view. More precisely, in Section $2$ we
summarize the basic aspects of \cite{Silverstein:2003hf} while at
the same time we investigate the equation of motions of this
tachyon like action for non flat FRW spacetimes. Also, an
asymptotic solution for the case of very small scale factors is
presented. Based on this, we illustrate a calculation of how to
obtain the probability of having the creation of a DBI closed
universe out of nothing at a certain distance away from the
throat. In Section $3$, there is a very systematic description of
the system described in Section $2$ in terms of the minisuperspace
formalism. Also, in a recent paper \cite{Lu:2005qy} a similar
analysis is performed using a different DBI action. We show, that
in some cases one can cast a ``time-tachyon"" dependent WDW
equation as it was first discussed in \cite{Sen:2002qa} and then
implemented in a stringy example \cite{Garcia-Compean:2005zn}. In
Section $4$, we introduce based on
\cite{Piao:2002nh,Chingangbam:2004ng}, several different types of
non-minimal couplings which are tachyon dependent and study the
WDW equation for the gravitational and tachyonic sectors. For a
particular choice of gravity-tachyon coupling we showed that the
minisuperspace description of cosmological moduli can be similar
to the minisuperspace description of certain interesting tachyon
based cosmological models. An interesting result is also recovered
according to which, the presence of the non-minimal coupling of a
particular form is absolutely necessary to render the
minisuperspace equation integrable in the case of a slowly varying
tachyon field. In Section $5$, we conclude our analysis by
providing some possible generalizations of the main results
presented in this article.

\section{DBI Action Of A Probe D$3$-brane In $AdS_5\times{S_5}$ Background}

In this section we review the key points of
\cite{Silverstein:2003hf} that will be the basis for our analysis.
We also follow the notation of this work. One of the reasons, that
D-brane dynamics is so rich in its structure, is due to the fact
that the action encompassing a D$3$-brane moving inside an ambient
$AdS_5\times{S_5}$ spacetime is the Dirac-Born-Infeld action. If
one is interested in studying small fluctuations around the radial
dimension of the AdS spacetime then the action takes the form
\cite{Aharony:1999ti}

\begin{equation}
S=-\frac{1}{g^2_{YM}}\int dtd^3x
f(\phi)^{-1}[-det(\eta_{\mu\nu}+f(\phi)\partial_{\mu}\phi\partial_{\nu}\phi)^{-\frac{1}{2}}-1]
\label{2}
\end{equation}
where the field $\phi=r/\alpha'$ is redefined in terms of the
radial direction $r$, the $\eta_{\mu\nu}=diag(-1,1,1,1)$ is the
flat background brane metric, while for the harmonic function we
choose

\begin{equation}
f(\phi)=\frac{\lambda}{\phi^4} \label{3}
\end{equation}

As is apparent, if we try to expand the radical in Eq.~(\ref{2}),
an infinite series of terms will emerge. In fact, all the terms of
the expansion are dependent on the velocity $\dot{\phi}^2$ of the
field $\phi$ and of its powers. It is then easy to see, that to
the lowest order in the velocities we recover a canonical kinetic
term for the radial motion while higher order ``velocities'' are
known to correspond to the presence of virtual W bosons at least
in the case of the time dependent field. Another interesting
observation of Eq.~(\ref{2}) is that after having expanded the DBI
action the constant potential term vanishes which is a
manifestation of the ``No-Force'' condition
\cite{Tseytlin:1996hi}. That can also be seen when computing the
energy of the system when $\phi=\phi(t)$

\begin{equation}
E=\frac{1}{\lambda}\phi^4(\frac{1}{\sqrt{1-\lambda\dot{\phi}^2/
\phi^4}}-1)\cong{\frac{1}{2}\dot{\phi}^2+\frac{1}{8}
\frac{\lambda\dot{\phi^4}}{\phi^4}}+... \label{4}
\end{equation}

The inclusion of antibranes in the system changes the picture
dramatically. The motion of $\bar{D3}$-brane probe in $AdS_5$ is
mainly dictated by the same equation Eq.~(\ref{2}) with one not so
innocuous alteration in the sign of the constant term in the
action

\begin{equation}
S=-\frac{N}{\lambda^2}\int
d^4x\phi^4(\sqrt{1-\lambda\dot{\phi}^2/\phi^4}+1) \label{5}
\end{equation}
where by N we denote the number of D-branes. This change in sign
leads to the presence of terms quartic in the field $\phi$ when
expanding in powers of $\lambda\dot{\phi}^2/\phi^4$, unlike in the
case of D$3$-brane.

The question one has to ask, is whether there is a way to
generalize Eq.~(\ref{2}) so as to study it in more general
 four dimensional backgrounds by possibly including extra potential
 terms. This can be achieved by direct covariantization of
 Eq.~(\ref{2}) down to four dimensions which now reads as follows

 \begin{equation}
 S=-\frac{1}{g_{YM}^2}\int
 d^4x\sqrt{-g}(f(\phi)^{-1}\sqrt{1+f(\phi)g^{\mu\nu}\partial_{\mu}\phi\partial_{\nu}\phi}+V(\phi)\mp{f(\phi)^{-1}})
 \label{6}
 \end{equation}
where the $\mp{}$ sign refers to the presence of D$3$-brane and
$\bar{D3}$-brane respectively, while $V(\phi)$ is an extra
potential which affects the cosmological evolution. Then by adding
the Einstein-Hilbert action for a general background $g_{\mu\nu}$

\begin{equation}
S_{E.H}=\int d^4x(\frac{1}{2}\sqrt{-g}M_{p}^2R+...) \label{7}
\end{equation}
one can safely study the cosmological equations of motions for the
system at hand.

One of the techniques implemented in \cite{Silverstein:2003hf} to
determine the cosmological evolution of the system under is
focused on writing down the FRW equations of state and attempting
to solve them. Thus, for a spatially flat line element

\begin{equation}
ds^2=-dt^2+a(t)^2dx^2 \label{8}
\end{equation}
one obtains through variation of the DBI action the expressions
for the energy density and the pressure  respectively,

\begin{equation}
\rho=\frac{\gamma}{f}+(V-f^{-1}) \label{9}
\end{equation}

\begin{equation}
p=-\frac{1}{\gamma f}-(V-f^{-1}) \label{10}
\end{equation}
where $\gamma$ defined by

\begin{equation}
\gamma=\frac{1}{\sqrt{1-f(\phi)\dot{\phi}^2}} \label{11}
\end{equation}
denotes the Lorentz factor for the moving brane. It is fairly
obvious then, that the positivity of the radical in $\gamma$ sets
an upper limit for the brane velocity in the bulk. Next, the
Einstein-Hilbert part of the action gives Friedman equations of
motion

\begin{equation}
3H^2=\frac{1}{g_sM_p^2}\rho \label{12}
\end{equation}

\begin{equation}
2\frac{\ddot{a}}{a}+H^2=-\frac{1}{g_sM_p^2}p \label{13}
\end{equation}
In addition, one has the field equation for the scalar field

\begin{equation}
\ddot{\phi}+\frac{3f'}{2f}(\dot{\phi})^2-\frac{f'}{f^2}
+\frac{3H}{\gamma^2}\dot{\phi}+
(V'+\frac{f'}{f^2})\frac{1}{\gamma^3}=0 \label{14}
\end{equation}
where the prime stands for differentiation with respect to $\phi$.

Solving this system of equations is not easy. Silverstein and Tong
though successfully solved the problem by resorting to the well
known ``Hamilton-Jacobi'' formalism \cite{Muslimov:1990be}.
Roughly speaking, by assuming that the Hubble parameter and the
potential can be expanded in terms of an infinite series of terms
with respect to $\phi$ and by proper manipulation of the
cosmological equations of motion they obtained analytical
solutions through iteration at the $\phi\rightarrow{0}$ limit. We
end this short review here by mentioning, that one of the
interesting solutions obtained through this technique is getting
quite exotic scale factors like $a(t)\sim{e^{-c/tM_p}}$, that may
be interpreted as a manifestation of the strong coupling between
gravity and the $\phi$ field. Also, it was shown, that after a
proper redefinition of the metric, the scalar field and the
potential of the action of this system can be brought to the usual
tachyon like DBI action. This observation will be crucial to our
analysis in translating our results, whenever possible, to tachyon
systems.

The analysis above stands only for flat spatial sections. It would
be interesting to explore other possibilities and topologies for
the universe as well. Hence, one of the most interesting examples
is the case of a closed universe with the $S^3\times{R}$ topology.
Studying this particular geometry is appealing also in terms of
quantum cosmology where one seeks to calculate the creation of a
close universe out of nothing. As we go along, we will show an
explicit formula based on certain approximations, for the
probability of the emergence of a D$3$ brane closed universe
through Linde's prescription \cite{Linde:1983mx}.

We begin with the closed FRW line element

\begin{equation}
ds^2=-dt^2+a(t)^2(d\psi^2+sin^2\psi(d\theta^2+sin^2\theta
d\varphi^2)) \label{15}
\end{equation}
The nontrivial spatial topology fashions the pressure and density
of the cosmic fluid so that

\begin{equation}
3H^2+\frac{3}{a^2}=\frac{1}{g_sM_p^2}\rho \label{16}
\end{equation}

\begin{equation}
2\frac{\ddot{a}}{a}+H^2+\frac{1}{a^2}=-\frac{1}{g_sM_p^2}p
\label{17}
\end{equation}
Obviously, Eq.~(\ref{12},\ref{13}) differ from
Eq.~(\ref{16},\ref{17}) by the addition of the $~1/a^2$ term. The
extra term though spoils the integrability of the model, therefore
it looks unlikely that a solution can be obtained through the
implementation of ``Hamilton-Jacobi'' method. However, there is a
way to circumvent this obstacle by observing that at very late
times the universe looks flat. Consequently, any significant
effects of the $S^3$ spatial topology will be more prevalent at
very small times, while at late times the system must behave like
\cite{Silverstein:2003hf}. Based on that, we seek cosmological
solutions for $t\rightarrow{0}$. Therefore, the aim of our
analysis is to obtain the asymptotic form for the scalar field,
the scale factor and the potential coupled to gravity. Towards
achieving our goal, we combine Eq.~(\ref{9},\ref{10},\ref{11})
with Eq.~(\ref{16},\ref{17}) into

\begin{eqnarray}
\sqrt{1-\frac{\lambda
\dot{\phi}^2}{\phi^4}}=\frac{-3\dot{\phi}^2}{g_sM_p^2(6H'\dot{\phi}-\frac{6}{a^2})}
=\frac{-3\dot{\phi}^2+\frac{2}{\lambda}\phi^4}{g_sM_p^2(6H'\dot{\phi}+6H^2)+\frac{2}{\lambda}\phi^4-V(\phi)}
\label{18}
\end{eqnarray}
Notice that Eq.~(\ref{18}) puts significant constraints on the
dynamics of the scalar field and in the geometry of the background
in general through both the upper limit for the $\dot{\phi}^2$ and
the positivity of the right hand side of the equalities.

 Let us consider the following ansatz for the scale factor,
field $\phi$ and  the potential

\begin{equation}
a(t)=ct^n \label{19}
\end{equation}
\begin{equation}
\phi(t)=\frac{\alpha}{t} \label{20}
\end{equation}
\begin{equation}
V(\phi)=\epsilon\phi^4 \label{21}
\end{equation}

It is quite clear now, that the model should be in a position to
determine the exact values of $(c,\alpha,n)$. It turns out, that
this particular ansatz is consistent with the cosmological
equations in the regime of small velocities
$(\lambda/\alpha^2<<1)$ by providing the numerical values of the
coefficients in Eq.~(\ref{19},\ref{20}). By substituting
Eq.~(\ref{19},\ref{20},\ref{21}) into Eq.~(\ref{18}) and keeping
terms that prevail in the asymptotic regime of $a\rightarrow{0}$
we obtain the following values for our parameters
$(\sqrt{2\epsilon g_sM^2_{P}/3},\sqrt{3/\epsilon},n=2)$. The
undetermined parameter $\epsilon$ is chosen to satisfy
$\lambda\epsilon<<1$. This condition is satisfied only for
$\epsilon<<1/\lambda$, if we demand the radius of the AdS
spacetime R to be much bigger than the string length since
$\lambda\sim{(\frac{R}{l_s})^4}$ \cite{Alishahiha:2004eh}. For
completeness we mention, that the equation of state for this
universe has the simple form

\begin{equation}
w\equiv{\frac{p}{\rho}}=-\frac{1}{3} \label{22}
\end{equation}
As a consistency check of our analysis, by substituting
Eq.~(\ref{16},\ref{17}) into Eq.~(\ref{22}) one can find regions
where the condition $\lambda\epsilon<<1$ is satisfied.

Our approach is focused on obtaining the scale factor of the
universe and the position of the moving brane (which is controlled
by the moduli $\phi$) as functions of time for a given potential
in the asymptotic regime of small scale factors. In essence, we
are mainly interested in getting a general idea of the dynamics of
the system rather than obtaining an exact solution which may not
even be possible due to the complexity of the FRW equations. More
precisely, our analysis, involves a series of truncations of terms
in Eq.~(\ref{18}) consistent with our regime. However, in this
region gravity effects are important and our analysis may not be
in a position to capture the totality of those effects that might
have been lost due to the truncation scheme we applied. Looking
back to Eq.~(\ref{20}) though, we see, that the brane moves away
form the throat as $t\rightarrow{0}$ at small velocities
($\lambda\varepsilon<<1$). Thus, we believe, that since the moduli
do not saturate the speed limit their dynamics is well described
by our model.

 The main motivation behind investigating the very early
evolution of this universe is to capture the curvature effects
that are involved to the non trivial nature of the $S^3$ spatial
topology for some particular approximations. Furthermore, since
the model is robust enough to provide us with the explicit
formulas of the scalar field, scale factor and potential, it
shouldn't be too hard to compute the probability for this universe
(which undergoes a power law expansion) to materialize to a given
size. Linde's approach will be the one implemented hereafter. To
this end, let us write down the total action decomposed as follows

\begin{equation}
S_{total}=S_{DBI}+S_{E.H} \label{23}
\end{equation}
where

\begin{equation}
S_{DBI}=-\frac{2\pi^2}{g_{YM}^2}\int
dt(\frac{a^3}{f(\phi)}\sqrt{1-f(\phi)\dot{\phi}^2}+a^3(V(\phi)-\frac{1}{f(\phi)}))
\label{24}
\end{equation}

\begin{equation}
S_{E.H}=6M_p^2\pi^2\int dt(-a\dot{a}^2+a) \label{25}
\end{equation}
Evidently, the Lagrangian inside the action depends only on the
temporal coordinate since the spacial integration is already
performed. Evaluating $S_{total}$ requires substituting
Eq.~(\ref{19},\ref{20},\ref{21}) into Eq.~(\ref{23}) as well as
performing the integration from $0$ to a finite but small time
$\tau$. We keep in mind, that $\tau$ must be quite small given the
fact that our whole analysis holds for the $t\rightarrow{0}$
regime.

By performing the integration on time (through the standard Wick
rotation) the action to the lowest order on $\tau$ shapes up as
follows

\begin{equation}
S_{total}=-\frac{16}{3}\pi^2M^3_P
g^{1/2}_s(\frac{2\epsilon}{3})^{1/2}\tau^3 \label{26}
\end{equation}
While the sign of the action has been a serious matter of
controversy at least for the case of simple gravity models we,
stick to Linde's formula for calculating the probability
$P=e^{-|S|}$.

Recasting Eq.~(\ref{26}) in terms of the scale factor $a_0$ (which
should be small enough according to our FRW solutions) that the
universe reaches just after its creation one gets
\begin{equation}
P=e^{-\frac{16}{3}\pi^2(\frac{3}{2\epsilon})^{1/4}M^{3/2}_P
g^{-1/4}_sa^{3/2}_0} \label{27}
\end{equation}
One observes that in Eq.~(\ref{27}) the 't Hooft coupling
$\lambda$ is missing, possibly depriving the system of its gauge
theory interpretation. However, we have to stress, that the
absence of $\lambda$ was a result of cancellations when we
expanded the square root in the DBI action and kept terms to the
lowest order in $\lambda$. Considering higher order terms in the
't Hooft coupling modifies Eq.~(\ref{27}) as follows

\begin{equation}
P=e^{-|\frac{16}{3}-\frac{\lambda\varepsilon}{18}+...|\pi^2(\frac{3}{2\epsilon})^{1/4}M^{3/2}_P
g^{-1/4}_sa^{3/2}_0} \label{28}
\end{equation}
All higher order terms in $\lambda$ appear in powers
$\lambda\varepsilon<<1$ and therefore have little effect on the
probability density other than slightly slowing down its decay
rate as the scale factor of the universe increases.

The formulae we obtained for the probability of the universe to
appear from nothing have a complementary interpretation. When we
computed the Euclidean action we integrated out not only the scale
factor but the moduli as well. Thus, Eq.~(\ref{27},\ref{28})
describe the probability of the universe to emerge out of nothing
having a small size and being at a large distance from the throat
according to Eq.~(\ref{19},\ref{20},\ref{21}).

What we described above is a direct implementation of one of the
tools developed through the years in the context of D$3$ physics.
It turns out, that we can get an analytical solution
Eq.~(\ref{27}) for a complicated system as the one described in
terms of the DBI action. In the following section, a different
pathway will be followed. The minisuperspace formalism will be
widely utilized in a very systematic way, so as to explore and
unfold the richness of the structure of stringy systems.

\section{Minisuperspace of the D$3$ Brane Action}

We start in the current section by surveying the fundamentals of
minisuperspace. WDW equation can be understood through the
canonical formalism of General Theory of Relativity first
developed by Arnowitt, Deser and Misner \cite{Arnowitt:1962hi}.
(For a nice exposition see \cite{Kolb:book}). According to the
Hamiltonian formalism of GR also known as ADM construction, the
spacetime is decomposed in $3+1$ in the following fashion through
the use of a lapse function $N$, a shift vector $N_{i}$ and an a
three dimensional spatial metric $h_{ij}$ (see \cite{Kolb:book}
for notation).

\begin{equation}
ds^2=g_{\mu\nu}dx^{\mu}dx^{\nu}=
N^2dt^2-h_{ij}((N^{i}dt+dx^{i})(N^{j}+dx^{j}) \label{29}
\end{equation}
To proceed further, we need to include the $S_{E.H}$ gravitational
action followed by a boundary term
\begin{equation}
S=-\frac{1}{16\pi G}\int d^4x\sqrt{g}(R(g)+2\Lambda)+\frac{1}{8
\pi G}\int_{\partial M} d^3x\sqrt{h}K \label{30}
\end{equation}
Then one can construct the Hamiltonian as follows

\begin{equation}
H=\int d^3x(\pi^{ij}\dot{h_{ij}}+\pi^{i}\dot{N^{i}}+\pi\dot{N}-L)
\label{31}
\end{equation}
where

\begin{equation}
\pi\equiv{\frac{\delta L}{\delta \dot{N}}}=0 \label{32}
\end{equation}

\begin{equation}
\pi^{i}\equiv{\frac{\delta L}{\delta \dot{N_{i}}}}=0 \label{33}
\end{equation}
The last two equations can be used to prove the independence of
the Hamiltonian density on the lapse function and the shift
vector. Therefore, it turns out, that the hamiltonian must vanish
identically. Based on this result the WDW equation comes out
naturally

\begin{equation}
H(\pi_{ij},h_{ij},\phi)\Psi(h_{ij},\phi)=0 \label{34}
\end{equation}
or equivalently
\begin{equation}
(\frac{G_{ijkl}}{(16\pi G)^2}\frac{\delta}{\delta
h_{ij}}\frac{\delta}{\delta
h_{kl}}+\frac{\sqrt{h}(R-2\Lambda)}{16\pi
G}-T^{0}_{0}(\phi,-i\frac{\partial}{\partial{\phi}}))\Psi(h_{ij},\phi)=0
\label{35}
\end{equation}
for a scalar field coupled to gravity and with $T^0_0$ energy
density.

Let us now implement the ADM construction into our system by
writing down the general ansatz for the line element

\begin{equation}
g_{\mu\nu}=-N^2dt^2+a^2(t)(d\psi^2+sin^2\psi(d\theta^2+sin^2\theta
d\varphi^2)) \label{36}
\end{equation}
Note that the symmetries of the FRW space give vanishing shift
vectors $N^{i}$.

The Enstein-Hilbert action reads
\begin{equation}
S_{E.H}=6M_p^2\pi^2\int
dt(\frac{a}{N^2}(\ddot{a}aN-a\dot{a}\dot{N}+\dot{a}^2N+N^3)
\label{37}
\end{equation}
By adding a total derivative term into the action or in other
words integration by parts will jettison the second derivative
with respect to time, and so we obtain

\begin{equation}
S_{E.H}=6M_p^2\pi^2\int dt(-\frac{a\dot{a}^2}{N}+aN) \label{38}
\end{equation}
Similarly the DBI action for the $D3-\bar{D3}$ system goes as
follows

\begin{equation}
S_{DBI}=-\frac{2\pi^2}{g^2_{YM}}(\int dt
(\frac{Na^3}{f(\phi)})\sqrt{1-\frac{f(\phi)}{N^2}\dot{\phi}^2}+Na^3(V(\phi)\mp{f^{-1}(\phi)}))
\label{39}
\end{equation}
Finally, the expression for the combined Lagrangian is

\begin{equation}
L=\gamma a N-\gamma\frac{a\dot{a}^2}{N}
-b\frac{Na^3}{f(\phi)}\sqrt{1-\frac{f(\phi)}{N^2}{\dot{\phi}^2}}
+bNa^3(-V(\phi)\pm{\frac{1}{f(\phi)}}) \label{40}
\end{equation}
where $b={2\pi^2}/{g^2_{YM}}$ and $\gamma=6M_{p}^2\pi^2$.

Next, in order to construct the Hamiltonian, the canonical momenta
for gravity and the scalar field need to be determined. In
addition to that, there is one extra constraint related to N that
should be used to give an identically vanishing Hamiltonian.
Setting the gauge $N=1$ before computing the momenta should be
avoided since it might give inconsistent results. Therefore, the
right way to go, is to keep track of the lapse function and only
at the end we should resort to the $N=1$ gauge. This is exactly
the procedure followed hereafter.

\begin{equation}
P_a\equiv{\frac{\partial{L}}{\partial{\dot{a}}}}=-\frac{2\gamma
a\dot{a}}{N} \label{41}
\end{equation}

\begin{equation}
P_{\phi}\equiv{\frac{\partial{L}}{\partial{\dot{\phi}}}}=
\frac{ba^3}{N}\frac{\dot{\phi}}{\sqrt{1-\frac{f(\phi)}{N^2}\dot{\phi}^2}}
\label{42}
\end{equation}
Therefore, the resulting Hamiltonian (though not yet in a
canonical form) is of the following form

\begin{equation}
H=-N(\gamma a+\frac{\gamma
a\dot{a}^2}{N^2}-\frac{ba^3}{f(\phi)}\sqrt{1-\frac{f(\phi)}{N^2}\dot{\phi}^2}
-\frac{ba^3}{N^2}\frac{\dot{\phi}^2}{\sqrt{1-\frac{f(\phi)}{N^2}}\dot{\phi}^2}+ba^3(-V(\phi)\pm{\frac{1}{f(\phi)}}))
\label{43}
\end{equation}

Finally one can verify, that the constraint
$\frac{\partial{L}}{\partial{N}}=0$ is such that substitution in
the Hamiltonian gives $H=0$. The last equation, is a direct
manifestation of diffeomorphism invariance of the theory. We have
now reached a point where it is safe to set $N=1$, which by itself
simplifies Eq.~(\ref{43}) quite a bit

\begin{equation}
H=-\gamma a-\gamma
a\dot{a}^2+\frac{ba^3}{f(\phi)}\frac{1}{\sqrt{1-f(\phi)\dot{\phi}^2}}
+ba^3(V(\phi)\mp{\frac{1}{f(\phi)}}) \label{44}
\end{equation}
Eq.~(\ref{41},\ref{42}), combined with Eq.~(\ref{44}) yield a
canonical expression for the Hamiltonian

\begin{equation}
H=-\gamma a-\frac{P_{a}^2}{4\gamma
a}+\frac{1}{f(\phi)}\sqrt{b^2a^6+P_{\phi}^2f(\phi)}+ba^3(V(\phi)\mp{\frac{1}{f(\phi)}})
\label{45}
\end{equation}

Before proceeding to the quantization of the system at hand, let
us focus on the radical in the Hamiltonian. Depending on whether
the scale factor overlaps in magnitude the contribution coming
from the momentum dependent term in the square root, we get
completely different quantum cosmology models. In fact, we shall
show, that the WDW equation becomes a linear differential equation
in the $\phi$ field for very small scale factors, instead of
second order. Also, we keep in mind, that the $P_{\phi}^2$ in the
radical remains positive according to Eq.~(\ref{42}), provided
that the $f(\phi)\dot{\phi}^2\leq{1}$ condition is satisfied at
all times. Moreover, although the spacetime covariance is spoiled
in the ADM decomposition the maximum speed limit for $\phi$
remains unaffected.

The first case to study is when $P_{\phi}^2f(\phi)<<b^2a^6$. In
this regime the Hamiltonians for the $D\bar{D}$ system are given
by

\begin{equation}
H_{D3}=-\gamma a -\frac{P_{a}^2}{4\gamma
a}+\frac{P_{\phi}^2}{2b^2a^3}+ba^3V(\phi) \label{46}
\end{equation}

\begin{equation}
H_{\bar{D3}}=-\gamma a -\frac{P_{a}^2}{4\gamma
a}+\frac{P_{\phi}^2}{2b^2a^3}+ba^3(V(\phi)+\frac{1}{2f(\phi)})
\label{47}
\end{equation}
When expanding the square root we only kept track of terms up to
quadratic order in $P_{\phi}$ neglecting the higher ones.
Therefore, this expansion is valid only for small values of the
canonical momentum. In addition, just to simplify things we set
$V(\phi)=0$. Then, canonical quantization enters by imposing
$P_{a}\rightarrow{-i\partial/\partial{a}}$,
$P_{\phi}\rightarrow{-i\partial/\partial{\phi}}$, which leads to

\begin{equation}
H\Psi(a,\phi)=
(-\frac{1}{2b^2a^3}\frac{\partial^2}{\partial{\phi}^2}
-\frac{1}{4\gamma
a^2}\frac{\partial}{\partial{a}}+\frac{1}{4\gamma
a}\frac{\partial^2}{\partial{a}^2}-\gamma a)\Psi(a,\phi)=0
\label{48}
\end{equation}
Note, that the inclusion of the extra derivative with respect to
the scale factor term which is introduced in Eq.~(\ref{48}) is due
to normal ordering upon quantization. It turns out that, the
differential equation we are dealing with can be solved through
separation of variables $\Psi(a,\phi)=\psi(a)\psi(\phi)$ which
gives

\begin{equation}
\frac{\partial^2{\psi(a)}}{\partial{a}^2}+\frac{1}{a}\frac{\partial{\psi(a)}}{\partial
a}+(-\epsilon_{1}a^2- \frac{\epsilon_{2}}{a^2})\psi(a)=0
\label{49}
\end{equation}

\begin{equation}
\frac{\partial^2{\psi(\phi)}}{\partial{{\phi}^2}}-2kb^2\psi(\phi)=0
\label{50}
\end{equation}
where $\epsilon_1=4\gamma^2$, $\epsilon_2=4\gamma k$ and k is a
constant. The exact form of the wave functions depends on the sign
of k. For $k>0$ we recover based on \cite{Tables of Integrals}

\begin{equation}
\psi(a)=Z_{\frac{\sqrt{\epsilon_2}}{2}}(\frac{\sqrt{\epsilon_1}}{2}a^2)
\label{51}
\end{equation}


\begin{equation}
\psi(\phi)=c_1e^{\sqrt{2kb^2}\phi}+c_{2}e^{-\sqrt{2kb^2}\phi}
\label{52}
\end{equation}
where $c_{1}$, $c_{2}$ are constants, and by Z we denote the
Bessel functions. The decaying solution tells us that the scalar
field will never acquire big VEV's simply because it's square
integrable wave function decays exponentially. In other words
$\phi$ prefers staying with a zero value (zero distance from the
AdS throat) since this is the value that the normalized wave
function maximizes the corresponding probability density. Keeping
the other solution translates to allowing the moduli to acquire
arbitrarily big values with even bigger {unbounded from above}
probability densities as $\phi$ grows bigger. Neglecting for the
moment the normal ordering factor, when k is negative and the
scale factor is quite large, the wave function of $\phi$ is
qualitatively different

\begin{equation}
\psi(a)=\sqrt{a}Z_{\frac{1}{4}}(\frac{i\sqrt{\epsilon_{1}}}{2}a^2)
\label{53}
\end{equation}


\begin{equation}
\psi(\phi)=c_1e^{i\sqrt{2|k|b^2}\phi}+c_{2}e^{-i\sqrt{2|k|b^2}\phi}
\label{54}
\end{equation}
In this case, $\psi(\phi)$ displays an oscillatory behavior,
however, the solution is not a pure phase allowing interference
between the modes. Later on, we will provide and analyze a
solution for the tachyon which is a pure phase.


Implementing the minisuperspace formalism in a useful way requires
imposing appropriate initial conditions so that the corresponding
wave functions have a proper interpretation. While in this
particular model it is not quite clear what should be a natural
set of initial conditions, we will provide an alternative way to
tell whether Eq.~(\ref{50}, \ref{51}) or Eq.~(\ref{52},\ref{54})
should be considered. Referring back to Eq.~(\ref{49}) the term
$\epsilon_{1}/a^2$ originates form the energy density of the
moduli in the low velocity regime, which goes as
$\rho=(1/2P_{\phi}^2b^2a^6)$. This is actually the energy density
of a scalar field. However the $\epsilon_{1}/a^2$ term can be
associated to the energy density of $\phi$ only when $k<0$. When
$k>0$ the above argument is no more applicable. Also, bare in mind
that Eq.~(\ref{54}) can be made square integrable for every value
of $\phi$ while the moduli wave function for $k>0$ diverges badly
when the brane starts moving away from the throat. On the other
hand, the gravity wave function while is not bounded from above
for negative $k$ as $a\rightarrow{\infty}$, it can be made square
integrable for sufficiently small $a$. Thus, the second set of
equations Eq.~(\ref{52},\ref{54}) seem to have a more concrete
quantum cosmology interpretation.



So far, our analysis was based upon the assumption that
$V(\phi)=0$, leaving some areas of the full theory unexplored.
However, one of the crucial limitations of minisuperspace
formalism is, that most of the times it is really difficult to get
analytical solution in closed form due to separability issues. To
evade this obstacle, we seek solutions where the scale factor is
very slow varying with respect to the scalar field and thus its
kinetic term can by ignored. Achieving this shouldn't be that hard
especially for universes that are around transition points between
phases of expansions and contractions. Therefore, this particular
requirement excludes the study of inflationary epochs. Hence, we
recover

\begin{equation}
(\frac{\partial^2}{\partial{\phi}^2}-\tilde{k}(V(\phi)+\frac{1}{2f(\phi)}))\psi_{\bar{D3}}(\phi)=0
\label{55}
\end{equation}

\begin{equation}
(\frac{\partial^2}{\partial{\phi}^2}-\tilde{k}V(\phi))\psi_{D3}(\phi)=0
\label{56}
\end{equation}
where $\tilde{k}=\frac{2b^3a^6}{2\lambda}$. For vanishing
potential the wave function for the D$3$-brane is linear in $\phi$
while for the $\bar{D3}$ is given by

\begin{equation}
\psi_{\bar{D3}}(\phi)=\sqrt{\phi}Z_{\frac{1}{6}}(\frac{i}{3}\sqrt{\tilde{k}}\phi^3)
\label{57}
\end{equation}

The qualitative differences between the two wave functions is not
a surprise to us. As can be seen from Eq.~(\ref{24}) the extra
$1/f(\phi)$ term in the Lagrangian of the anti-brane attracts the
scalar field by hindering it from moving away from the throat at
early times while this doesn't occur with the D$3$-brane due to
the no force condition. To be more specific, let us focus on
investigating the case of a quartic potential of the following
type $V=V_0\phi^4$. It turns out, that both wave functions have
similar form

\begin{equation}
\psi_{\bar{D3}}(\phi)=
\sqrt{\phi}Z_{\frac{1}{6}}(\frac{i}{3}\sqrt{\tilde{k}
(V_0+\frac{1}{2\lambda})}\phi^3) \label{58}
\end{equation}

\begin{equation}
\psi_{D3}(\phi)= \sqrt{\phi}Z_{\frac{1}{6}}
(\frac{i}{3}\sqrt{\tilde{k}(V_0)}\phi^3) \label{59}
\end{equation}
While Eq.~(\ref{58},\ref{59}) increase linearly in $\phi$ in the
asymptotic regime of $\phi\rightarrow{0}$, the probability density
corresponding to $\psi_{\bar{D3}}$ is bigger in magnitude than
$|\psi_{D3}|^2$, due to the presence of the $1/2\lambda$ term.
This can be attributed to the fact, that the $\bar{D3}$- brane
lingers longer around the throat due to the stronger attraction it
feels compared to the moving D$3$-brane.

 Referring back to Eq.~(\ref{45}) one observes that in the case
of very small scale factors (possibly early times) or in the case
where $P^2_{\phi}f(\phi)>>b^2a^6$, or in both cases where
$f(\phi)\dot{\phi}^2\leq{1}$ the Hamiltonian will be

\begin{equation}
H=-\gamma a-\frac{P^2_{a}}{4\gamma
a}+f^{-1/2}(\phi)P_{\phi}+ba^3(V(\phi)\mp{\frac{1}{f(\phi)}})
\label{60}
\end{equation}
The last term can be taken to be equal to zero. This is justified
in the context of the approximation we just described, but it can
also be accomplished by choosing $ V(\phi)=\frac{1}{f(\phi)}$ for
D$3$-branes, while there is no positive potential $V(\phi)$ one
can choose to make the last term in Eq.~(\ref{60}) vanish. Either
way, the WDW equation reads

\begin{equation}
(-\gamma a +\frac{1}{4\gamma a}\frac{\partial^2}{\partial{a}^2}
-if^{-1/2}\frac{\partial}{\partial{\phi}})\Psi(a,\Phi)=0
\label{61}
\end{equation}
while the corresponding wave functions to the lowest order of $a$
are

\begin{equation}
\psi(\phi)=c_1e^{\frac{ik\lambda^{1/2}}{\phi}} \label{62}
\end{equation}

\begin{equation}
\psi(a)=c_2A_{i}(\sqrt{4k\gamma}a)+c_3B_{i}(\sqrt{4k\gamma}a)
\label{63}
\end{equation}
where k is a constant coming from the separation of variables,
$c_1,c_2,c_3$ are integration constants and $A_{i}, B_{i}$ are the
Airy functions.

There is a remarkable property of Eq.~(\ref{61}) which we now
describe. It is unambiguously a first order differential equation
with respect to $\phi$. Additionally, it is shown
\cite{Silverstein:2003hf}, that the late time behavior of $\phi$
for a quartic potential is $\phi\sim{\sqrt\frac{\lambda}{t}}$. So,
based on the fact that there is a well established correspondence
between $(\phi\leftrightarrow{T})$ given by
$T=\sqrt{\frac{\lambda}{\phi}}$ \cite{Silverstein:2003hf}, and on
the result that $T\sim{t}$ as $t\rightarrow{\infty}$,
Eq.~(\ref{61}) describes a time dependent WDW equation
\cite{Sen:2002qa},\cite{Garcia-Compean:2005zn}.

Regarding $\psi(\phi)$, we see that it is a pure phase exhibiting
an extreme oscillatory behavior between ($-1,1$) in the vertical
axis for $c_1=1$ when approaching closer to the AdS throat, while
it reaches unity at infinite distance away from the origin. This
particular result holds for both $D3$ and $\bar{D3}$ and it might
have to do with strong gravity-scalar interaction.

As far as $\psi(a)$ is concerned, it turns out, that in order to
satisfy DeWitt boundary condition $(\psi(a=0)=0)$ the
gravitational wave function must be of the form \cite{Handbook}

\begin{equation}
\psi(a)\sim{{\sum^{\infty}_{0}
3^k(\frac{2}{3})_{k}\frac{(\sqrt{4k\gamma}a)^{3k+1}}{{(3k+1)!}}}}
\label{64}
\end{equation}

\begin{figure}
\includegraphics{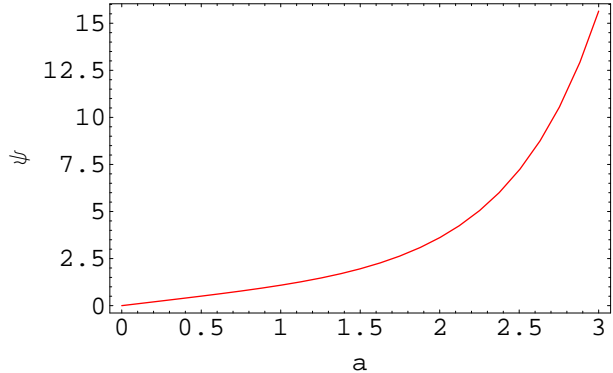}
\caption{\label{fig1}Plot of the $\psi(a)$ wave function that
comes about as a solution of Eq.~(\ref{61}), as a function of the
scale factor of the universe, in the regime where a is very small
and $4k\gamma=1$. Initial conditions used $\psi(a=0)=0,
\frac{d}{da}\psi(a=0)=1$.}
\end{figure}

We conclude, by providing the graph of $\psi(a)$ as a function of
the scale factor for $4k\gamma=1$. This shows, that the universe
emerges out of nothing (zero scale factor) from zero probability
density to a nonzero scale factors with $|\psi(a)|^2>0$. We note,
that in general satisfying the Dewitt boundary condition is really
hard since in most cases it leads to trivial models (i.e.
vanishing wave function for any scale factor). However, in our
analysis no such problem is encountered.

\section{Quantum Cosmology for Non-minimal Tachyon-Gravity
couplings}

The DBI action associated to the moduli field $\phi$ is known to
be similar to the one of the tachyon. This is easily seen for a
tachyon potential of the form $\sim{T^{-4}}$ followed by a proper
redifintion of fields \cite{Silverstein:2003hf}. Even though the
model we described in the previous section has no relation to the
tachyon other than the fact that they both share similar actions
(since Silverstein-Tong model is shown \cite{Silverstein:2003hf}
to be free of tachyonic instabilities) it would be interesting to
see whether one can uncover other similarities in their dynamics
when viewed from a quantum cosmology perspective. We expect for
instance, that for a typical tachyonic potential (which is widely
used in tachyon cosmology) of the following form
$V(T)\sim{e^{-T^2}}$, the cosmological equation of motions of the
tachyon will differ in general from the corresponding FRW
equations of the moduli $\phi$ due to the non trivial potential.
Things can get more complicated if higher order gravity tachyonic
couplings are included. Thus, in this section our primary concern
is to employ the minisuperspace formalism for the case of the
tachyon being non-minimally coupled to gravity. While such
couplings have been well explored \cite{Sen:1999md} in terms of
string theory, there hasn't been any specific quantum cosmological
application so far. We also note, that in
\cite{Piao:2002nh},\cite{Chingangbam:2004ng} the authors
investigated the importance of this type of coupling inside the
scope of inflationary cosmology. We basically follow the notation
of \cite{Chingangbam:2004ng} through the rest of the section.

Let us begin by considering the following action describing the
non-minimal tachyon-gravity coupling

\begin{equation}
S=\int d^4x\sqrt{-g}(
\frac{M^2_{P}}{2}Rf(T)-AV(T)\sqrt{1+Bg^{\mu\nu}\partial_{\mu}T\partial{\nu}T})
\label{65}
\end{equation}
where $A=\sqrt{2}/(2\pi)^3g{\alpha'}^2$, $B=8ln2{\alpha'}$,
$M^2_{P}=u/g^2{\alpha'}$ and $u=2V_{6}/(2\pi)^7{\alpha'}^3$.

With $V_6$ we denote the volume of the compactified manifold while
u is a dimensional constant. The tachyon potential $V(T)$ can be
taken to have the following asymptotic behavior:
$V(T)\sim{e^{-T^2}}$ for $T\rightarrow{0}$ while for
$T\rightarrow{\infty}$ the potential reads $V(T)\sim{e^{-T}}$.
These are the conventional forms used in tachyon cosmology. While
in our system B is a constant, one can consider it to be a
function of the tachyon field as well \cite{Choudhury:2002xu}. The
action we investigate is not written in the Einstein frame.
Therefore, it is wise to perform the following tachyon dependent
conformal transformation $g_{\mu\nu}\rightarrow{f(T)g_{\mu\nu}}$
that transforms Eq.~(\ref{65}) into

\begin{equation}
S=\int
d^x\sqrt{g}(\frac{M^2_{P}}{2}(R-\frac{3}{2}\frac{f'^2}{f^2}g^{\mu\nu}\partial_{\mu}T\partial_{\nu}T
)-A\tilde{V(T)}\sqrt{1+Bf(T)g^{\mu\nu}\partial_{\mu}T\partial{\nu}T})
\label{66}
\end{equation}
where the rescaled tachyon potential reads
$\tilde{V(T)}=V(T)/f(T)^2$. For a closed FRW metric the Lagrangian
is as follows

\begin{equation}
L=\frac{6\pi^2}{M^2_{P}}((a(1-\dot{a}^2)+\frac{1}{4}\frac{f'^2}{f^2}a^3\dot{T}^2)M^4_{P})-\frac{M^2_{p}}{6}A\tilde{V(T)}a^3\sqrt{1-Bf\dot{T}^2})
\label{67}
\end{equation}
For the case where the tachyon velocity is much smaller than its
upper limit, namely $1>>Bf\dot{T}^2$, then by expanding the
radical up to first order in $\dot{T}^2$ we end up with

\begin{equation}
L=\frac{6\pi^2}{M^2_{P}}(a(1-\dot{a}^2)M^4_{P}-\frac{M^2_{P}}{6}A\tilde{V}a^3+(\frac{1}{4}M^2_{P}\frac{f'^2}{f^2}+\frac{M^2_{P}}{12}AB\tilde{V}f)a^3\dot{T}^2)
\label{68}
\end{equation}

Next, we repeat the same step by step procedure implemented in the
former section in constructing the Hamiltonian of WDW equation
corresponding to Eq.~(\ref{68}). The associated canonical momenta
are of the form

\begin{equation}
P_a=-12\pi^2M^2_{P}a\dot{a} \label{69}
\end{equation}

\begin{equation}
P_{T}=\frac{12\pi^2}{M^2_{P}}(\frac{1}{4}M^4_P\frac{f'^2}{f^2}+\frac{M^2_{P}}{12}AB\tilde{V(T)}f)a^3\dot{T}
\label{70}
\end{equation}
Thus the Hamiltonian defined as $H=P_{a}\dot{a}+P_{T}\dot{T}-L$
reads

\begin{equation}
H=M^{-4}_P(-\frac{M^2_{P}}{24\pi^2}\frac{1}{a}P^2_{a}+\frac{M^2_{P}}{24\pi^2}\frac{1}{a^3}\frac{P^2_{T}}{(\frac{1}{4}(\frac{f'}{f})^2
+\frac{1}{12M^2_{P}}AB\tilde{V}f)} +6\pi^2
M^6_{P}(-a+\frac{A\tilde{V}}{6M^2_P}a^3)) \label{71}
\end{equation}
Proceeding with the canonical quantization of the Hamiltonian at
hand and by implementing the ansatz $\Psi(a,T)=\psi(a)\psi(T)$ for
the wave function, we find

\begin{eqnarray}
\frac{M^2_P}{24\pi^2}\frac{1}{a\psi(a)}\frac{\partial^2\psi(a)}{\partial{a}^2}+\frac{M^2_P}{24\pi^2}\frac{1}{a^2\psi(a)}\frac{\partial\psi(a)}{\partial{a}}\nonumber
-\frac{M^2_P}{24\pi^2}\frac{1}{a^3(\frac{1}{4}(\frac{f'}{f})^2+\frac{AB\tilde{V}f}{M^2_P})}\frac{1}{\psi(T)}\frac{\partial^2\psi(T)}{\partial{T^2}}\\
 +6\pi^2M^6_P(-a+\frac{A\tilde{V}a^3}{6M^2_P})=0 \label{72}
\end{eqnarray}

One of the reasons Eq.~(\ref{72}) is a bit cumbersome, is that the
non-minimal coupling gives additional contributions to the kinetic
term of the rolling tachyon. Thus, it is quite reasonable to
expect a lack of integrability in the equation. As we proceed we
will mainly focus on two particular interesting cases. For the
first one, our motivation is to obtain cosmological wave functions
for both gravity and the tachyon for $\tilde{V}$ and $f$ that
satisfy the slow roll condition for inflation to occur
\cite{Chingangbam:2004ng} (even though fitting with the
observational data might be difficult). For the second one, we
attempt through a particular choice of coupling to render WDW
equation integrable and consequently obtain the corresponding wave
functions.

Working in the regime where $a>>\frac{A\tilde{V}a^3}{6M^2_P}$ the
resulting equations are

\begin{equation}
\frac{M^2_P}{24\psi^2}\frac{1}{(\frac{1}{4}(\frac{f'}{f})^2+\frac{AB\tilde{V}}{12M^2_P})}
\frac{1}{\psi(T)}\frac{\partial^2\psi(T)}{\partial{T^2}}=k
\label{73}
\end{equation}

\begin{equation}
\frac{M^2_P}{24\pi^2}\frac{a^2}{\psi(a)}\frac{\partial^2\psi(a)}{\partial{a}^2}
+\frac{M^2_P}{24\pi^2}\frac{a}{\psi(a)}\frac{\partial\psi(a)}{\partial{a}}
-6\pi^2M^6_Pa^4-k=0 \label{74}
\end{equation}
Then by choosing

\begin{equation}
\tilde{V(T)}=1-\frac{1}{4}{T^4}+... \label{75}
\end{equation}

\begin{equation}
f(T)=1-\frac{1}{2}T^2+\frac{1}{4}T^4 \label{76}
\end{equation}
together with the condition $AB=6M^2_P$ \cite{Chingangbam:2004ng}
while keeping terms to the lowest order in T, one gets

\begin{equation}
\psi(T)=c_1e^{-\sqrt{2\delta_2}T}+c_2e^{\sqrt{2\delta_2}T}
\label{77}
\end{equation}

\begin{equation}
\psi(a)=Z_{\frac{\sqrt{\delta_2}}{2}}(\frac{\sqrt{\delta_1}}{2}a^2)
\label{78}
\end{equation}
where, $\delta_1=144\pi^4k$, $\delta_2=12\pi^2k/M^2_P$, and k is
an integration constant which is taken to be positive. It should
be kept in mind, that this solution is valid only close to the top
of the tachyon potential ($V\sim{e^{-T^2}}$), where inflation can
take place. The form of the tachyon wave function is surprisingly
simple despite the presence of non-minimal coupling. One can also
prove, that the presence of the exponentially increasing term in
the $\psi(T)$ is not a problem since the $\psi(T)$ can be shown to
be square integrable for small values of the tachyon and with a
particular choice of the constants involved.

There is a striking similarity between Eq.~(\ref{51},\ref{52}) and
Eq.~(\ref{77},\ref{78}). Apart from trivial differences in the
constants involved, they all appear to be essentially identical.
This is quite surprising considering the inclusion of nontrivial
couplings in the WDW Hamiltonian. Thus, when the tachyon is very
close to the top of the potential the minisuperspace description
of the tachyon in a closed four dimensional universe is the same
as Silverstein-Tong's model (for a brane of $S^3$ spatially
topology). Also, this particular correspondence holds for $k<0$ as
one can very easily show. As we shall show later on, this is no
longer true when the tachyon starts rolling away from the top i.e.
$V(T)\sim{e^{-T/2}}$ for certain types of non minimal gravity
tachyon couplings.

In both regimes of small and big values for T, one can find
non-minimal couplings that render WDW equation completely
integrable. This is easy to see by inspection of Eq.~(\ref{72}).
Thus, by demanding that $\tilde{V(T)}=\tilde{V_0}=const$, when
$V(T)=e^{-T}$ and the size of the universe is small we obtain

\begin{equation}
\psi(T)=Z_{4\sqrt{\lambda_1}}(4\sqrt{\lambda_{2}}e^{-T/4})
\label{79}
\end{equation}

\begin{equation}
\psi(a)=Z_{\frac{\sqrt{\xi_2}}{2}}(\frac{\sqrt{\xi_1}}{2}a^2)
\label{80}
\end{equation}
where $\lambda_1=3\pi^2k/2M^2_P$, $\lambda_2=2\pi^2kABV _0$ and
$\xi_1=144\pi^4M_P$, $\xi_2=24\pi^2k/M^2_P$. Both formulae are
obtained for $f(T)=e^{-T/2}/\sqrt{\tilde{V_0}}$ and for positive
integration constant k.

For the case where $V(T)=e^{-T^2}$, the appropriate conformal
coupling that maintains integrability is
$f(T)=e^{-T^2/2}/\sqrt{\tilde{V_0}}$. As a result, the WDW
equation for the tachyon reads

\begin{equation}
\frac{\partial^2{\psi(T)}}{\partial{T^2}}=(\frac{1}{4}\eta_1T^2+\eta_1\eta_2e^{-T^2/2})\psi(T)
\label{81}
\end{equation}
where $\eta_1=24\pi^2k/M^2_P$, $\eta_2=AB\tilde{V_0}/12M^2_P$, and
$k>0$.

\begin{figure}
\includegraphics{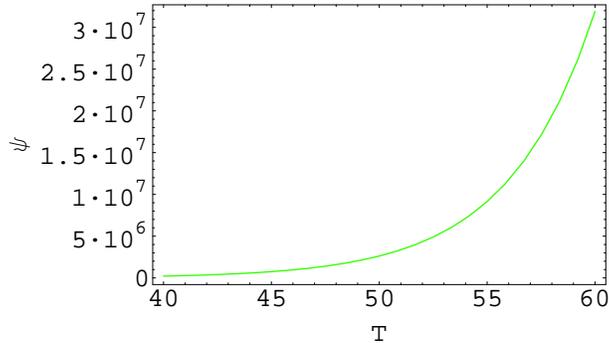}
\caption{\label{fig4} Tachyonic wave function Eq.~(\ref{79})
plotted for when $24\pi^2k=M^2_P$ and $ AB\tilde{V_0}=12M^2_P$.}
\end{figure}

We solve  Eq.~(\ref{81}) numerically and show, that even in the
case where the tachyon is close to the top of the potential
$V(T)=e^{-T^2}$ its wave function is increasing almost linearly
for small T. Similarly, when the tachyon is away from the top i.e.
$V=e^{-T}$ its wave function increases quite rapidly as a function
of T as $T\rightarrow{\infty}$. The fact that the wave function is
not square integrable, prevents us from interpreting our results
in terms of probability densities. However, in theories with
tachyons, we expect that as the tachyon rolls it has to reach
infinity namely $T\rightarrow{\infty}$, as the condensation takes
place. There is no other option for it but to increase its
magnitude. Therefore, we expect the ``probability density'' to
find the tachyon in higher expectation values has to increase
constantly and maybe that is the reason for unbounded solutions to
the WDW Equation.

\begin{figure}
\includegraphics{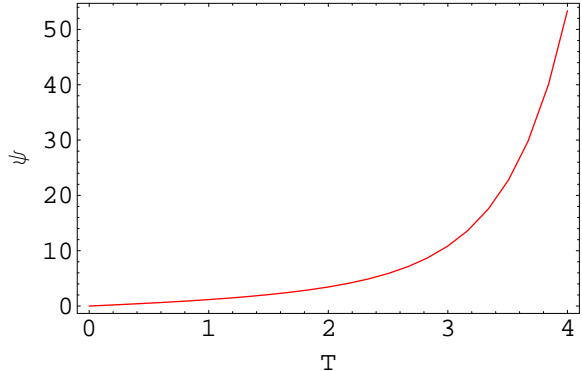}
\caption{\label{fig2}Plot of the tachyon wave function $\psi(T)$
that solves Eq.~(\ref{81}), when $24\pi^2k=M^2_P$ and $
AB\tilde{V_0}=12M^2_P$.}
\end{figure}

\section{Conclusions}

In this paper we present several quantum cosmology aspects of
gravitational theories tied to the Dirac-Born-Infeld Lagrangian.
Based on $AdS/CFT$ duality, in the strong coupling limit of $N=4$
super Yang Mills theory, the dynamics of the moduli field $\phi$
can be described in terms of the DBI action for a probe D$3$-brane
moving in AdS spacetime. Then, we seek cosmological solutions for
a universe with $S^3$ spatial topology described by the DBI
action. We were able to find the asymptotic expressions for the
scale factor and the scalar field for a given external potential
coupled to gravity. A formal expression for the creation of this
particular universe out of ``nothing'' to a finite small scale
factor is recovered in terms of Linde's prescription.

The above system was further analyzed in terms of the
minisuperspace formalism and WDW equation. A variety of solutions
was presented for the scale factor and the moduli wave function,
both being qualitatively different. At a certain limit, it is
shown, that the wave function of the moduli (which become first
order differential equation in T) is similar to the one of the
tachyon presented in previous works resulting in a time dependent
version of the WDW after a proper redefinition of variables. In
the later part of this article, there was a discussion regarding
the effects of a non-minimal gravity coupling with the tachyon
field. In particular, we showed that one can introduce
phenomenologically interesting couplings for which the tachyon
have similar minisuperspace description with the moduli of D-brane
inspired models \cite{Silverstein:2003hf}. Thus, our analysis
clearly indicates the importance of non minimal gravity-tachyon
couplings at least within the framework of quantum cosmology. It
was also shown, that even for such highly coupled systems, a
particular choice of conformal coupling can make the WDW equation
of the tachyon to decouple from gravity when the size of the
universe is quite small.

One of the possible extensions of the current work would be to
implement our minisuperspace description to other possible ambient
topologies as was done in \cite{Linde:2004nz} (yielding regular
solutions for zero scalar factors) and generalized in
\cite{McInnes:2005su} for a toroidal universe in terms of
Hartle-Hawking and Firouzjahi-Sarangi-Tye wave function
\cite{Firouzjahi:2004mx}. Moreover, further studies of quantum
cosmology on compact, flat tachyonic (unstable) universes might
also address the issue of accelerating singular toral universes
\cite{McInnes:2005sa}. Finally, in one of the non-minimal
couplings that we implemented to make WDW equation integrable i.e.
$f(T)=e^{-T^2/2}/\sqrt{\tilde{V_0}}$, it would be interesting to
search for appropriate values for the constant $\tilde{V_0}$ for
which inflation could occur. Towards this goal, one may introduce
an external tachyon potential in the action and study its effects
on the universe, even beyond the minisuperspace formalism.

 \noindent
{\bf Acknowledgements}\\

 The author would like to thank Ioannis Bakas for his constant
support, motivation and for his comments on the manuscript. I also
thank Pran Nath, Yogi Srivastava, Toamsz Taylor and Allan Widom
for carefully reading and commenting on the paper. Finally, i
would like to thank the organizers of ``Corfu Summer Institute on
EPP 2005'' for their hospitality, ample financial support and for
the excellent, friendly and stimulating environment which
contributed in finalizing this work.

\end{document}